\documentclass{aa}
\usepackage{graphicx}
\begin{document}
\title{Variable stars in nearby galaxies}
\subtitle{V. Search for Cepheids in Field A of NGC 6822
\thanks{Based on observations collected at ESO-La Silla}}
\author{E. Antonello, D. Fugazza, L. Mantegazza, M. Stefanon, S. Covino}
\institute{Osservatorio Astronomico di Brera, Via E.~Bianchi 46,
       I--23807 Merate, Italy}
\offprints{E. Antonello} 
\date{Received 20 December 2001 / Accepted 20 february 2002}
\titlerunning{Field A of NGC 6822}
\authorrunning{E. Antonello et al.}

\abstract
{The results of a CCD survey for variability of stars in the nearby 
galaxy NGC 6822 are presented. The goal of the survey was to obtain good 
light curves of Cepheids for Fourier decomposition and to detect shorter 
period Cepheids. Since the program was carried 
out with a relatively small telescope, the Dutch 0.9 m at ESO-La Silla, 
the observations were unfiltered (white light, or $Wh$--band). 
The analysis revealed the presence of more than 130 variable stars. 
21 population I Cepheids are detected; 6 of them were  
already known from previous works (Kayser, \cite{kay}). For at least three 
Cepheids, however, the previous identification or period was wrong.
Some probable population II (W Vir) stars are also identified.
The dispersion of the fundamental mode Cepheid $PL$ relation appears 
to be small.
}
\maketitle

\section{Introduction}
Cepheids are variable stars that are used to measure distances of
galaxies in the Local Group and nearby clusters (e.g. Madore et al. 
\cite{m1}), and are the primary calibrator for the secondary standard 
candles that are applied at much greater distances (e.g. Jacoby et al. 
\cite{ja}). However, they are not only fundamental stars as primary 
distance indicators, but are also an essential tool for testing the 
theories on the internal constitution of stars and stellar evolution. 
There are several problems yet to be solved. The radiative codes used to
construct nonlinear pulsation models proved to be unable to agree 
with observations when applied to the comparison of Cepheid characteristics in
the Galaxy and in Magellanic Clouds (e.g. Buchler \cite{buc}). 
Resonances among the pulsation modes give rise to observable effects on 
the light curves which can be exploited to put constraints on the pulsational 
models and on the mass-luminosity relations. When these resonances 
observed in Cepheids of the Galaxy and Magellanic Clouds are used to constrain 
purely radiative models, one obtains stellar masses that are too small 
to be in agreement with stellar evolution calculations. According to Buchler 
et al. (\cite{bu2}), it is clear that some form of convective 
transport and of turbulent dissipation is needed to make progress; this has 
been proved for example for the first overtone mode Cepheids in the Galaxy
(Feuchtinger et al. \cite{feu}).

The MACHO, EROS and OGLE projects dedicated to the detection of microlensing 
events in the direction of the Magellanic Clouds produced enormous amount of 
data on variable stars in these galaxies (e.g. Welch et al. \cite{we}; 
Beaulieu \& Sasselov \cite{bs}; Udalski et al. \cite{uda}). More recently, 
the project DIRECT was dedicated to the massive CCD photometry of M31 and M33 
with the purpose of detecting Cepheid and eclipsing binaries for direct 
distance determination of these galaxies (e.g. Kaluzny et al. \cite{kal};
Macri et al. \cite{mac}). The purpose of our project was to obtain good light 
curves of Cepheids to extend the comparison of the characteristics of 
these stars in different galaxies. In order to exploit the telescope time 
and reach the faintest luminosities, our strategy was to observe in white 
light, i.e. without filter. Massive CCD photometry of the nearby 
galaxy IC 1613 has been already discussed in the previous papers of this
series (Mantegazza et al. \cite{man}, and references therein; see also the 
recent survey by Udalski et al. \cite{ud2}), while in this paper 
we present the first results for NGC 6822. 
\begin{figure*}
\resizebox{\hsize}{!}{\includegraphics{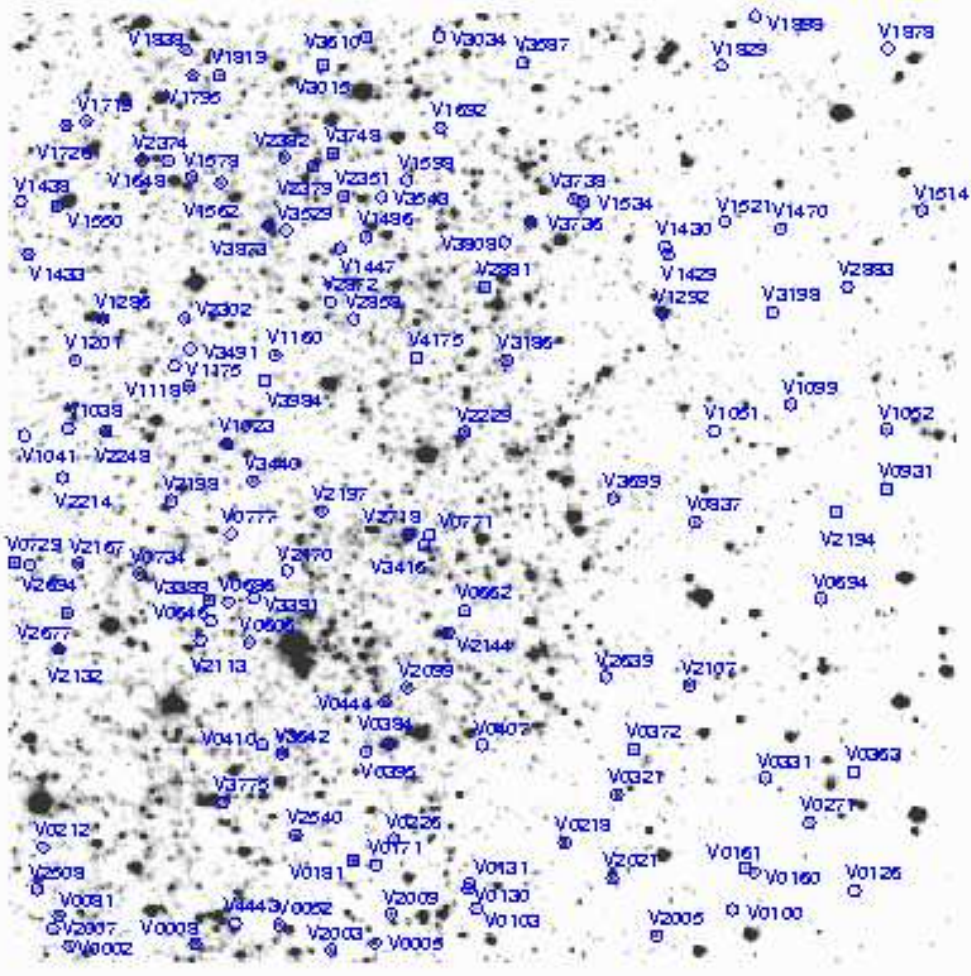}}
\caption[ ]{Field A in NGC 6822 with the detected variable stars;
north is up and east is to the right
}  
\label{mappa}
\end{figure*}
\begin{table*}
\caption[]{Cepheids in Field A of NGC 6822}
\begin{flushleft}
\begin{tabular}{llllllll}
\hline\noalign{\smallskip}
Name & $\alpha$(2000) & $\delta$(2000) & P & $Wh$ & $V$-$R$ & mode/type & \\
     &  [$^h$~~$^m$~~$^s$]  & [$^o$~~ '~~ ''] & [d] &  &  &   &      \\
\noalign{\smallskip}\hline\noalign{\smallskip}
V0052 & 19 44 55.50 &  -14 50 05.3 &   8.3288 &  20.70 &  0.69 & F & \\
V0171 & 19 44 57.14 &  -14 49 50.1 &   3.6500 &  21.82 &  1.25 & F & \\
V0271 & 19 45 04.57 &  -14 49 38.0 &   6.0220 &  21.11 &  0.83 & F & \\
V0372 & 19 45 01.54 &  -14 49 20.5 &   6.7361 &  20.99 &  0.99 & F & \\
V0410 & 19 44 55.15 &  -14 49 20.8 &   3.9993 &  21.22 &  0.80 & F & \\
V0444 & 19 44 57.24 &  -14 49 09.7 &   3.5946 &  20.18 &  0.67 & 1O? & \\
V0686 & 19 44 54.53 &  -14 48 45.6 &   5.2889 &  21.28 &  1.32 & F & \\
V0729 & 19 44 50.85 &  -14 48 36.5 &   9.3647 &  20.40 &  1.00 & F & V25\\
V0734 & 19 44 53.00 &  -14 48 38.9 &  16.82   &  19.75 &  1.07 & F & V9\\
V0771 & 19 44 57.97 &  -14 48 28.0 &  16.74   &  22.02 &  0.96 & W Vir \\
V1286 & 19 44 52.31 &  -14 47 35.9 &  41.92   &  19.81 &  0.70 & Ceph? \\
V1292 & 19 45 01.90 &  -14 47 32.3 & 122.7    &  17.72 &  0.85 & F & V13\\
V1433 & 19 44 51.01 &  -14 47 20.2 &   6.8586 &  20.81 &  0.46 & F & \\
V1447 & 19 44 56.36 &  -14 47 17.4 &   3.0504 &  22.02 &  0.97 & F & \\
V2113 & 19 44 54.07 &  -14 48 55.2 &   6.15   &  22.77 &  0.78 & W Vir \\
V2374 & 19 44 52.93 &  -14 46 56.3 &  29.21   &  19.13 &  1.14 & F & \\
V2382 & 19 44 55.37 &  -14 46 55.1 &   3.8831 &  21.06 &  0.78 & F & V3\\
V2858 & 19 44 56.61 &  -14 47 34.9 &   5.483  &  22.82 &  0.69 & W Vir \\
V3015 & 19 44 56.01 &  -14 46 32.2 &   5.4927 &  21.06 &  0.76 & F & \\
V3186 & 19 44 59.25 &  -14 47 44.6 &   4.5561 &  21.05 &  1.09 & F & \\
V3529 & 19 44 55.43 &  -14 47 13.2 &   1.7112 &  21.25 &  0.81 & 1O? & \\
V3543 & 19 44 57.06 &  -14 47 04.7 &  22.1    &  21.90 &  1.00 & W Vir \\
V3736 & 19 44 59.62 &  -14 47 10.3 &  30.499  &  18.94 &  0.93 & F & V1\\
V3738 & 19 45 00.36 &  -14 47 04.4 &   8.9464 &  20.70 &  1.20 & F & \\
V3873 & 19 44 55.12 &  -14 47 12.2 &  37.52   &  18.36 &  0.78 & F & V2\\
V3984 & 19 44 55.10 &  -14 47 50.5 &   1.4230 &  21.84 &  0.68 & 1O? & \\
\noalign{\smallskip}
\hline
\end{tabular}
\end{flushleft}
\end{table*}

\section{Observations of NGC 6822}
The irregular galaxy NGC 6822 [$\alpha(2000)=19^h44^m56^s$, $\delta(2000)= 
-14\degr 48' 02''$, l=25\degr, b=-18\degr], was studied by Hubble
(\cite{hub}, who discovered 11 Cepheids and other bright irregular
variables, and then surveyed by Kayser (\cite{kay}) using photographic 
plate material obtained by Arp and Baade at the prime focus of the 
Palomar 5m telescope. She found 13 Cepheids, with period $P$ in the 
range bewteen 10 and 90 d. More recently, CCD light curves were 
obtained for 6 known Cepheids by Schmidt \& Spear (\cite{sch}). 
NGC 6822 is close to the galactic plane, resulting in large foreground 
reddening. A short review of the problems related to the period-luminosity
$PL$ relation and distance modulus estimates of this galaxy was given by 
Madore \& Freedman (\cite{mf}).

Our observations were performed with the direct CCD camera attached to the 
Dutch 0.91m telescope of the La Silla Astronomical Observatory (ESO) 
during 3 runs from 1996 to 1998. The available CCD 
detector was the ESO chip No. 33, which is a TEK CCD with 512x512
pixels, pixel size of $27{\mu}m$ and spatial resolution of 0{\farcs}44,
providing a field of view of 3{\farcm}77x3{\farcm}77. Given the limited size 
of the field of view, the need to observe not too far from the meridian and at 
the same time to be able to get two images of the same field on the same night,
we were forced to limit our programme to a few selected fields of NGC 6822.
Most of the observations were performed without filter (white light, 
hereinafter $Wh$) in order to get the best photon statistics for the study 
of faint Cepheid light curves. The properties of this photometric band 
are discussed in Paper I (Antonello et al. \cite{an2}) and Paper III 
(Antonello et al. \cite{an3}) of this series; other information can 
be obtained from the paper by Riess et al. (\cite{rie}).
Moreover, two images were taken in Johnson $V$ and $R$ filter to obtain 
information on star colors.
In this paper we present the results regarding Field A which
contains the largest number of previously--known variable 
stars. 

\section{Data Reduction, Calibration and Analysis}
\begin{figure}
\resizebox{\hsize}{!}{\includegraphics{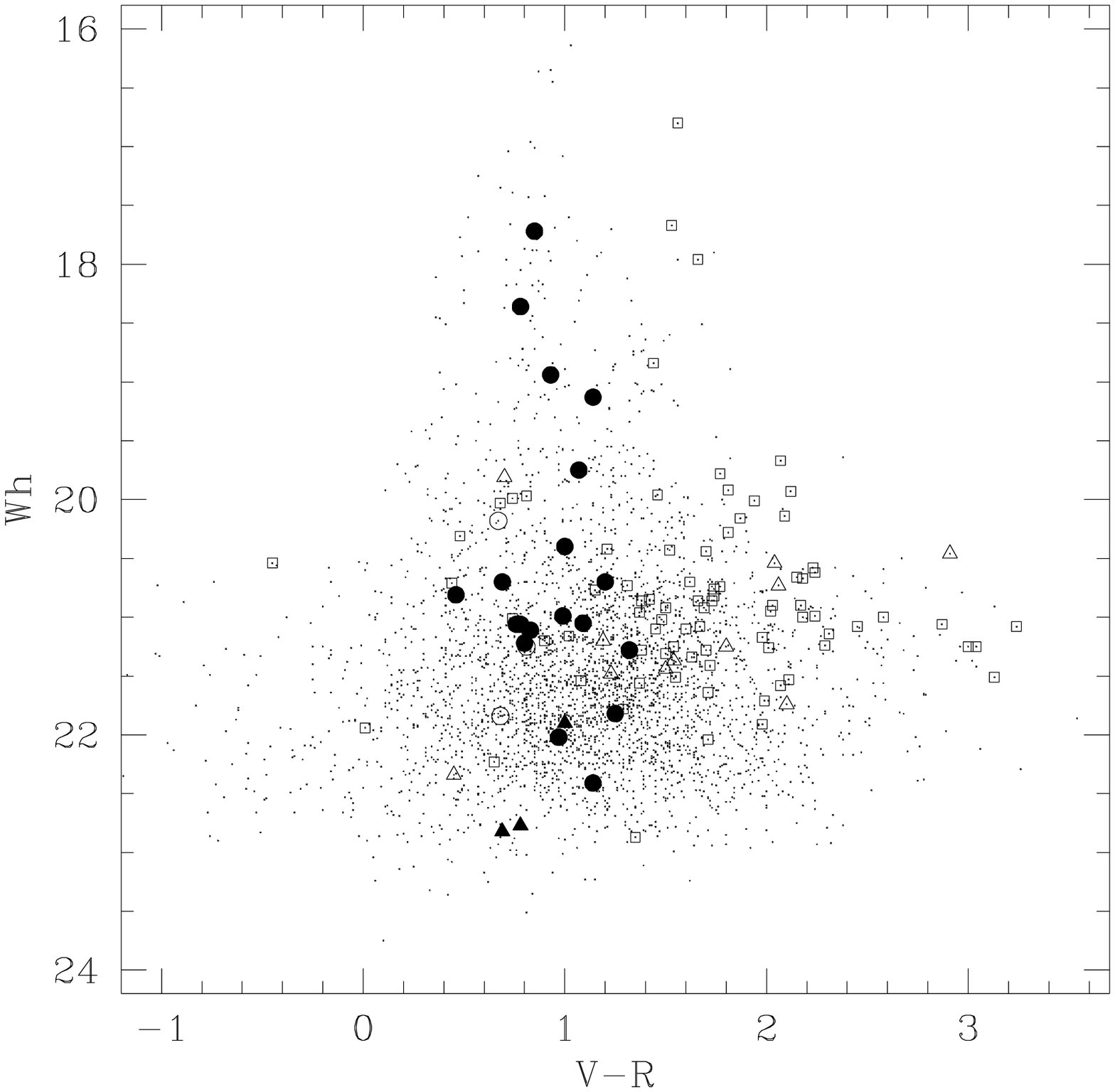}}
\caption[ ]{Color-magnitude ($V$--$R$,$Wh$) diagram of stars in Field A.
{\em Filled circles:} fundamental mode Cepheids; 
{\em open circles:} probable first overtone mode Cepheids;
{\em filled triangles:} probable population II Cepheids (W Vir); 
{\em open triangles:} other periodic variables;
{\em squares:} other semiregular, irregular and possible long $P$ variables
}  
\label{color}
\end{figure}

The methods of data reduction and calibrations were the same as those 
discussed in Paper III. The instrumental $wh$ magnitudes were 
transformed into our ``standard'' $Wh$ system  using 150 stars 
with at least 31 measurements; all the variable candidates were
excluded. We found:
\begin{equation}
V-wh=-1.28-0.04(V-R)+0.38(V-R)^2
\end{equation}
 and hence
\begin{equation}
Wh=wh-1.28 
\end{equation}
The search for variable stars was performed with the various techniques
described in Paper I and III in order to minimize the probability of
wrong identifications. 
In addition, we made the following test. By means of 
numerical simulations we generated 10000 time-series of random numbers 
normally distributed for each set of observations (from 25 to 34 data points),
we analyzed them with the least squares power spectra for different 
frequency intervals of investigation, and computed the distributions 
of the amplitudes of the highest peak in each spectrum.  
Among the stars selected as variable candidates with the previous methods, 
we confirmed only those whose highest peak in the least-squares power 
spectrum had a probability less than 0.01 of being generated by random noise.
As a result of the application of these techniques we detected 130 variable
stars, i.e. 21 Cepheids, 18 other periodic variables and 91 irregular 
or semiregular variables. Previously in this field only 6 Cepheids and 3 
long period variables were known (Kayser, \cite{kay}).
The location of the variables in the field is shown in Fig. \ref{mappa};
the color-magnitude diagram is shown in Fig. \ref{color}. 

\section{Results}
\begin{figure*}
\resizebox{\hsize}{!}{\includegraphics{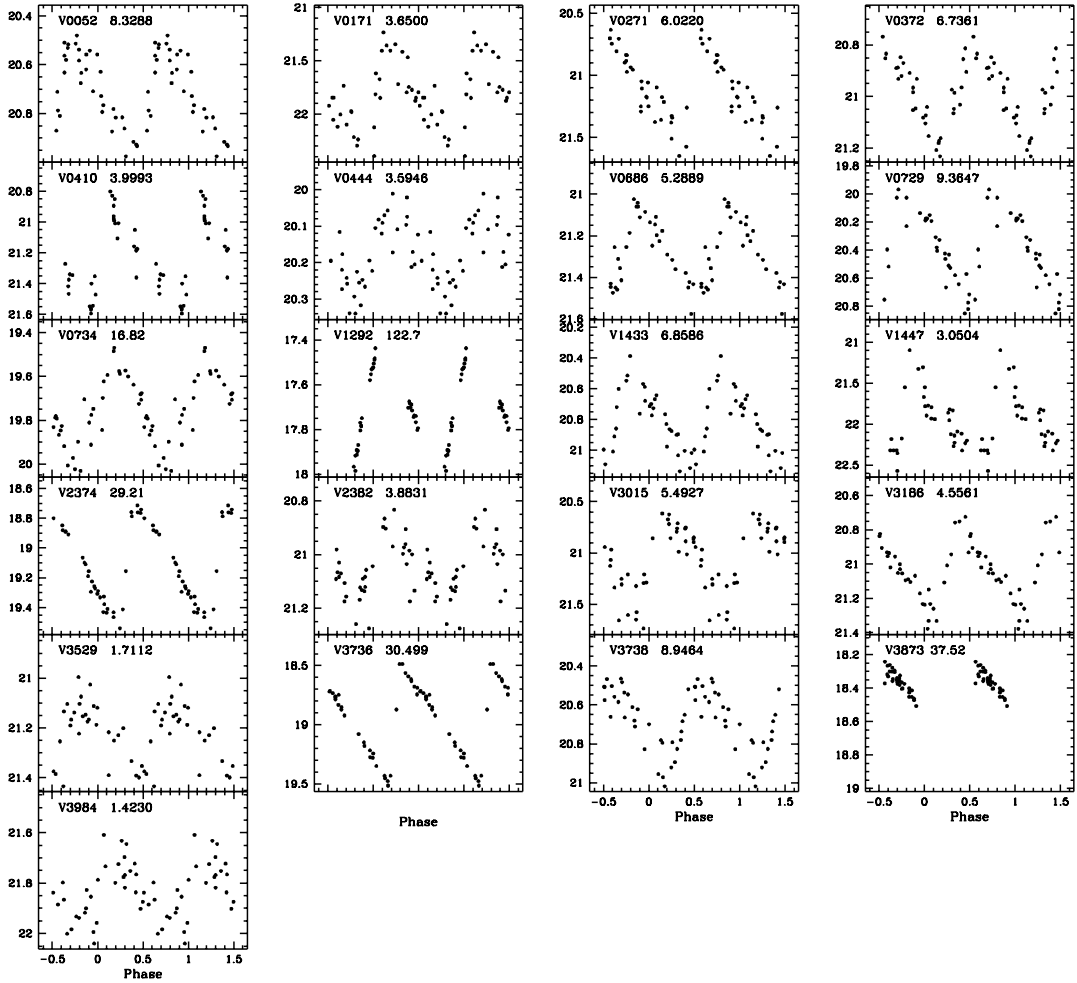}}
\caption[ ]{Cepheid $Wh$ light curves. For each star, the identification
number and the period are reported}  
\label{light1}
\end{figure*}

Cepheid variable stars are listed in Table 1, along with their equatorial 
coordinates, the period $P$, the mean $Wh$ magnitude, the random
phase $V-R$ index, the pulsation mode or stellar type, and the previous
identification number (Kayser, \cite{kay}). 
The light curves phased according to 
the periods of Table 1 are shown in Fig. \ref{light1}. It should be noted
that, given the data distribution, the determination of the right period 
for values longer than about 20 d is not easy, since a 
number of peaks in the power spectrum supply equivalent fits.
Some comments on a few objects follow.

{\em V0729}: Kayser classified this star as an irregular variable, without 
ruling out a possible $P$ of about 13 d. Our data show unambiguously that 
the star is periodic with $P=9.364$ d.

{\em V0734}: Hubble (1925) classified it as a Cepheid, but according to Kayser 
it should be a rapid irregular variable. Our data support Hubble's conclusion 
and give a period of 16.8 d, which is in agreement with the value of 16.9 d 
supplied by him.

{\em V1292}: this is the longest period Cepheid of our sample. 
Our data set indicate $P \sim 123$ d while Kayser obtained $P=90$ d.
As a check, we analyzed her data and found that the best--fitting $P$ is 
actually about 120.7 d; in the power spectrum the peak corresponding to 90 d
is much lower. By merging our and the Kayser datasets, shifting them to 
a common zero--point and rescaling the amplitude, we obtained a best--fitting 
$P$ of 120.66 d. The light curve probably does not repeat
exactly from cycle to cycle; this characteristic is common among the longer
$P$ Cepheids both in the Milky Way and Magellanic Clouds. The long $P$ value
places the star beyond the limit of about 100 d which is usually assumed 
for the validity of the $PL$ relation (in metal--poor galaxies); owing to the 
nonlinear pulsation cycle characteristics (e.g. Aikawa \& Antonello 
\cite{aa}), the $PL$ relation in this $P$ range tends to have a negligible
slope.

{\em V3873}: also in this case our (marginally) best $P$, 25.35 d differs 
from Kayser's one, 37.44 d. However, the analysis of the merged dataset 
supports 37.44 d as the best $P$.

Other periodic variables have been detected and four of them are probable 
W Vir stars. V1286 ($P$=41 d), which has an intermediate colour, 
is too bright to be a W Vir, and is probably too faint to be a Cepheid. 
The list of all the other periodic variables and of 91 long $P$ 
and irregular ones are available upon request from the first author.
\begin{figure}
\resizebox{\hsize}{!}{\includegraphics{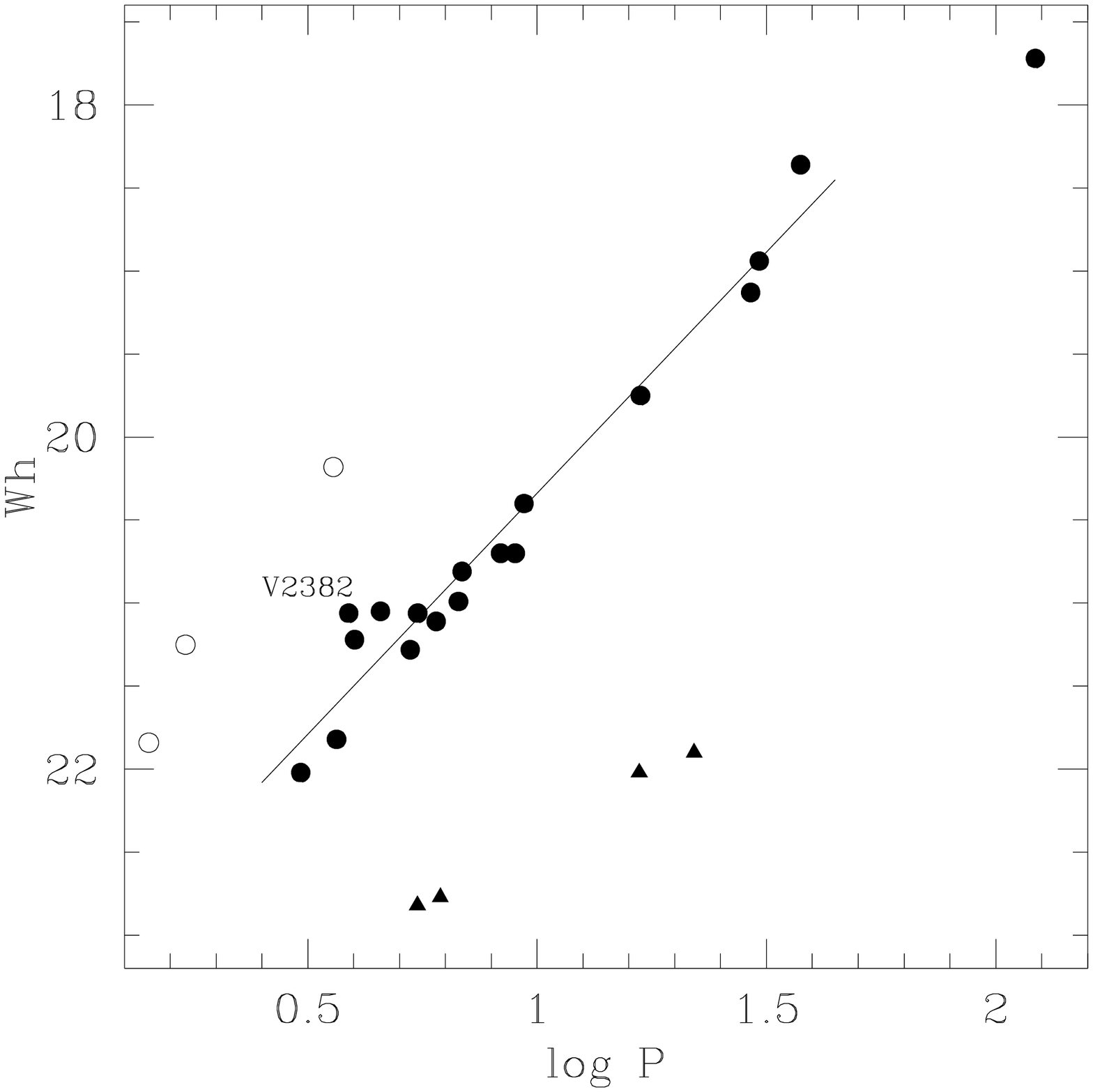}}
\caption[ ]{$PL$ diagram for Cepheids in Field A of NGC 6822.
{\em Filled circles:} fundamental mode Cepheids; {\em open circles:} probable
first overtone mode Cepheids; {\em filled triangles:} probable population
II Cepheids. The line is the statistical relation 
obtained for fundamental mode Cepheids; for these stars 
$V \sim Wh+0.3$, typically
}
\label{pl}
\end{figure}
\section {Conclusion}
We have presented the first results of a survey of NGC 6822 for
detecting and studying Cepheids; since a relatively small telescope was used,
the observations were performed without a filter. Fainter Cepheids than 
$V \sim 22$ were found. In the analogous case of IC 1613 we were 
able to detect Cepheids as faint as $V \sim 23$; the better performances are
probably due to the unique characteristics of IC 1613, i.e. a very
low background and less severe problems given by crowding, and to the large
number of observations (more than 60 data points per star).
In NGC 6822, 21 population I Cepheids were identified, while only 6 of them 
were previously known in the same field. Of these six Cepheids, V0729 and
V0734 were previously classified as irregular variable by Kayser (\cite{kay};
Hubble however indicated V0734 as a Cepheid), and V1292 has a much 
longer $P$ than previously reported. 
The $PL$ diagram is shown in Fig. \ref{pl}. It is possible to derive a 
$PL$ relation for Cepheids in the $Wh$ band, with a similar slope to that 
which could be obtained for $V$ and $R$ data; the slope is --2.89 and the 
zero-point is 23.23. The relation appears to be narrow. We note in particular 
that the star
V2382 has the lowest amplitude ($\Delta{Wh}=0.29$), therefore we could suspect
it is a binary, and the true luminosity of the Cepheid should be fainter.
As in the case of IC 1613, we have performed a few observations in the 
$BVRI$ bands with the WFI at the 2.2 m ESO telescope, in order to apply the 
method devised by Freedman (\cite{fre}) for deriving the $PL$ relation in 
the various bands of the standard photometry, using the $Wh$ light curves 
(e.g. Antonello et al. \cite{an2.5}).
Presently the $Wh$ observations of NGC 6822 are almost completed, and in the 
next papers we will report on the other observed fields, discuss the 
Cepheid light curve characteristics and make a comparison with the stars
of other galaxies in the Local Group; moreover we will present the results 
of the standard photometry $PL$ relations.

\begin{acknowledgements}
Part of this work was supported by MURST--COFIN, project 9802844774-005.
\end{acknowledgements}

\end{document}